\documentclass[showpacs,preprintnumbers,prd,nofootinbib,floats,amssymb,floatfix]{revtex4}
\usepackage{amsmath}
\usepackage{amsfonts}
\usepackage{graphicx}
\usepackage{hyperref}
\usepackage{amsmath}
\usepackage{amstext}
 
\begin{document}

\title { Hamiltonian Dynamics for an alternative action describing Maxwell's equations }
\author{ Alberto Escalante }  \email{aescalan@ifuap.buap.mx}
\affiliation{ Instituto de F{\'i}sica Luis Rivera Terrazas, Benem\'erita Universidad Aut\'onoma de Puebla, (IFUAP). Apartado
 postal J-48 72570 Puebla. Pue., M\'exico,}
 \author{ Omar Rodr{\'i}guez Tzompantzi}  
 \affiliation{ Facultad de Ciencias F\'{\i}sico Matem\'{a}ticas, Benem\'erita Universidad Au\-t\'o\-no\-ma de Puebla,
 Apartado postal 1152, 72001 Puebla, Pue., M\'exico.}

\begin{abstract} 
 We develop a complete  Dirac's canonical analysis  for  an alternative  action  that   yields   Maxwell's four-dimensional equations of motion. We study in detail the full symmetries of the action  by following all steps of Dirac's method  in order to obtain  a detailed description of symmetries.  Our results indicate that  such an  action does not have  the same symmetries than Maxwell theory,  namely, the  model  is not a gauge theory and the number of physical degrees of freedom are different. 
\end{abstract}

\date{\today}
\preprint{}
\maketitle
\section{INTRODUCTION}
\vspace{1em} \
A dynamical system is characterized by means of its symmetries which  constitute  an  important information  in both the  classical and quantum context.  The physics of the fundamental interactions  based on  the standard model \cite{1},   is a relevant example where the symmetries of a dynamical system  just like gauge covariance, CPT invariance, the identification of the physical degrees of freedom and conserved quantities are useful for understanding  the classical and  quantum formulation  of the  theory.  However, the Lagrangians studied in the standard model  are singular systems and the conventional Hamiltonian analysis,  is not the correct way for studying them \cite{2}. On the other hand, it is well-known that the analysis of
a dynamical system by means of its equations of motion  implies that the phase space is not endowed with a natural or preferred symplectic structure as it  has been claimed in \cite{3,4}, and the freedom in the choice of the symplectic
structure is an important issue because could yield different quantum formulations. Hence, in spite of we have an infinity ways to choose a symplectic structure for any system, the next question arises: are   the symmetries of the classical theory preserved in all  different choices of  the  symplectic structure?,   the answer in general  is not \cite{5,6}. In fact, we will show along  this paper that the alternative action to conventional Maxwell's  equations  proposed in  \cite{7,8}  is not a gauge theory, is not invariant under parity  symmetry and the physical degrees of freedom are not those knew for  the electromagnetic field. Thus,  in the study of the symmetries of a dynamical system must be taken into account an action principle plus their equations of motion, because    the action gives  the equations of motion and additionally  fixes  the symplectic structure of the theory  \cite{4}.  \\
The  alternative action proposed in \cite{7,8}   is a singular system, and this fact was ignored in those works. Nevertheless, we shall perform our study in a different way, this is, we will develop a complete  Dirac's canonical  approach. This  formalism is an elegant approach for obtaining the relevant physical information of a theory under study, namely, the counting of physical degrees of freedom, the gauge transformations, the study of the constraints, the extended Hamiltonian and the extended action \cite{2}, being a relevant information because is the guideline to make the best progress in the analysis of quantum aspects of the theory. All those facts will be explained along the paper. 
 \newline
\newline
\section{Hamiltonian Dynamics for an alternative action describing Maxwell's equations  }
The system  that we shall  study in this paper is given by  the following action principle \cite{7,8}
\begin{equation}
S[E_{i},B_{i}]=\frac{1}{8\pi}\int \left[B^{i}\left (\frac{1}{c}\partial_{t}E_{i}-\epsilon{_{i}}^{jk}\partial_{j}B_{k}\right)-E^{i} \left(\frac{1}{c}\partial_{t}B_{i}+\epsilon{_{i}}^{jk}\partial_{j}E_{k} \right)\right]dx^4,
\label{eq.1}
\end{equation}
where  $E$  and  $B$  represents the electric  and magnetic fields and  form a set of six dynamical variables.\\
By considering to $E_{i}$ and $B_{j}$ as our set of dynamical  variables, the equations of motion obtained from the action (\ref{eq.1})  are given by 
\begin{eqnarray}
\begin{cases}
\begin{array}{c}
\partial_{t}E_{i}=c\epsilon_{i}{^{jk}}\partial_{j}B_{k},\\
\partial_{t}B_{i}=-c\epsilon_{i}{^{jk}}\partial_{j}E_{k},
\end{array}\end{cases}
\end{eqnarray}
which correspond to the half of  Maxwell's  equations. In order to obtain the complete Maxwell equations without sources, we observe that the fields should satisfy   
\begin{eqnarray}
\nabla\cdot  E&=&0, \nonumber  \\ 
\nabla \cdot B&=&0, 
\label{eq.3}
\end{eqnarray}
 in our analysis we shall take into account the above equations. In \cite{7,8} the action (\ref{eq.1}) was  proposed as an alternative Lagrangian for describing the dynamics of the electromagnetic field, however, it is easy to see that  the action is  neither Lorentz invariant nor invariant under parity symmetry;  this fact will be reflected in the hamiltonian analysis,  in particular in the number of physical degrees of freedom of the theory.   Furthermore,  we will show below that the action does not have  the principal symmetry  knew for Maxwell theory namely, gauge invariance.  \\ 
For our aims,  we identify from the action principle that the  Hessian given by 
\[
H^{ij}=\frac{\partial^{2}\mathcal{L}}{\partial(\partial_{t}E_{i})\partial(\partial_{t}B_{j})}
\]
has entries zero, so the system  under study is a singular theory. The Hessian   has  a  rank=0,  and  six null vectors. Therefore we expect six primary constraints.  A pure Dirac's analysis calls  for the definition of the momenta $(\Pi^{i}_{E}, P^{i}_{M})$  canonically conjugate to ($E_{i}$, $B_{i}$)
\begin{equation}
\Pi_{E}^{i}=\frac{\partial\mathcal{L}}{\partial(\partial_{t}E_{i})}=\frac{1}{8\pi c}B^{i},
\end{equation}
\begin{equation}
P_{M}^{i}=\frac{\partial\mathcal{L}}{\partial(\partial_{t}B_{i})}=-\frac{1}{8\pi c}E^{i}.
\end{equation}
where the following six primary constraints arise 
\begin{eqnarray}
\Phi{}_{E}^{i}&:&\Pi_{E}^{i}-\frac{1}{8\pi c}B^{i}  \approx 0, \nonumber \\
\Psi_{M}^{j}&:&P_{M}^{j}+\frac{1}{8\pi c}E^{j} \approx 0.
\label{eq.8}
\end{eqnarray}
In order to verify that the above constraints are  the correct,    we calculate the Jacobian among them
\[
\frac{\partial(\Phi{}_{E}^{i},\Psi_{M}^{j})}{\partial(E_{i},B_{j},\Pi_{E}^{i},P_{M}^{j})}=\left(\begin{array}{cccccccccccc}
0 & 0 & 0 & -\frac{1}{8\pi c} & 0 & 0 & 1 & 0 & 0 & 0 & 0 & 0\\
0 & 0 & 0 & 0 & -\frac{1}{8\pi c} & 0 & 0 & 1 & 0 & 0 & 0 & 0\\
0 & 0 & 0 & 0 & 0 & -\frac{1}{8\pi c} & 0 & 0 & 1 & 0 & 0 & 0\\
\frac{1}{8\pi c} & 0 & 0 & 0 & 0 & 0 & 0 & 0 & 0 & 1 & 0 & 0\\
0 & \frac{1}{8\pi c} & 0 & 0 & 0 & 0 & 0 & 0 & 0 & 0 & 1 & 0\\
0 & 0 & \frac{1}{8\pi c} & 0 & 0 & 0 & 0 & 0 & 0 & 0 & 0 & 1
\end{array}\right),
\]
we observe that the Jacobian has  rank=6, and it  is constant on the constraints  surface. Therefore, the primary constraints given in (\ref{eq.8})  are the set of correct   primary constraints.\\
Now, the canonical  Hamiltonian takes the form
\begin{eqnarray}
H_{c}&\equiv&\int\mathcal{H}_{c}d^{3}x=\int[\dot{E}_{i}\Pi_{E}^{i}+\dot{B}_{i}P_{M}^{i}-\mathcal{L}]d^{3}x\\\nonumber
&=&\int[\dot{E}_{i}(\frac{1}{8\pi c}B^{i})+\dot{B_{i}}(-\frac{1}{8\pi c}E^{i})-\frac{1}{8\pi c}B_{i}\partial_{t}E_{i}+\frac{1}{8\pi c} B_{i}\epsilon_{ijk}\partial_{j}B_{k}\\\nonumber
&+&\frac{1}{8\pi c}E_{i}\partial_{t}B_{i}+ \frac{1}{8\pi c}E_{i}\epsilon_{ijk}\partial_{j}E_{k}]d^{3}x\\\nonumber
&=&\frac{1}{8\pi}\int[B_{i}\epsilon_{ijk}\partial_{j}B_{k}+E_{i}\epsilon_{ijk}\partial_{j}E_{k}]d^{3}x.\\\nonumber
\end{eqnarray}
Hence,  the primary Hamiltonian is given by 
\[
H^{1}=H_{c}+\int u_{i}(x)\Phi{}^{i}(x)d^{3}x,
\]
where  $u_{i}$ are Lagrange multipliers enforcing the primary constraints.\\
In order to know  if there are secondary constraints, we calculate the following matrix whose entries are the Poisson brackets among the primary constraints, this is  
\[
\begin{cases}
\begin{array}{c}
\left\{ \Phi{}_{E}^{i}(x),\Phi{}_{E}^{j}(y)\right\} =0, \\
\left\{ \Psi{}_{M}^{i}(x),\Psi{}_{M}^{j}(y)\right\} =0, \\
\left\{ \Phi{}_{E}^{i}(x),\Psi{}_{M}^{j}(y)\right\}  =-\frac{1}{4\pi c}\delta^{ij}\delta(x-y),
\end{array}\end{cases}
\]
So, the matrix acquires the form 
{\scriptsize
\[
W^{\mu\nu}=\left(\begin{array}{cccccc}
\left\{ \Phi{}_{E}^{1}(x),\Phi{}_{E}^{1}(y)\right\}  & \left\{ \Phi{}_{E}^{1}(x),\Phi{}_{E}^{2}(y)\right\}  & \left\{ \Phi{}_{E}^{1}(x),\Phi{}_{E}^{3}(y)\right\}  & \left\{ \Phi{}_{E}^{1}(x),\Psi{}_{M}^{1}(y)\right\}  & \left\{ \Phi{}_{E}^{1}(x),\Psi{}_{M}^{2}(y)\right\}  & \left\{ \Phi{}_{E}^{1}(x),\Psi{}_{M}^{3}(y)\right\} \\
\left\{ \Phi{}_{E}^{2}(x),\Phi{}_{E}^{1}(y)\right\}  & \left\{ \Phi{}_{E}^{2}(x),\Phi{}_{E}^{2}(y)\right\}  & \left\{ \Phi{}_{E}^{2}(x),\Phi{}_{E}^{3}(y)\right\}  & \left\{ \Phi{}_{E}^{2}(x),\Psi{}_{M}^{1}(y)\right\}  & \left\{ \Phi{}_{E}^{2}(x),\Psi{}_{M}^{2}(y)\right\}  & \left\{ \Phi{}_{E}^{2}(x),\Psi{}_{M}^{3}(y)\right\} \\
\left\{ \Phi{}_{E}^{3}(x),\Phi{}_{E}^{1}(y)\right\}  & \left\{ \Phi{}_{E}^{3}(x),\Phi{}_{E}^{2}(y)\right\}  & \left\{ \Phi{}_{E}^{3}(x),\Phi{}_{E}^{3}(y)\right\}  & \left\{ \Phi{}_{E}^{3}(x),\Psi{}_{M}^{1}(y)\right\}  & \left\{ \Phi{}_{E}^{3}(x),\Psi{}_{M}^{2}(y)\right\}  & \left\{ \Phi{}_{E}^{3}(x),\Psi{}_{M}^{3}(y)\right\} \\
\left\{ \Psi{}_{M}^{1}(x),\Phi{}_{E}^{1}(y)\right\}  & \left\{ \Psi{}_{M}^{1}(x),\Phi{}_{E}^{2}(y)\right\}  & \left\{ \Psi{}_{M}^{1}(x),\Phi{}_{E}^{3}(y)\right\}  & \left\{ \Psi{}_{M}^{1}(x),\Psi{}_{M}^{1}(y)\right\}  & \left\{ \Psi{}_{M}^{1}(x),\Psi{}_{M}^{2}(y)\right\}  & \left\{ \Psi{}_{M}^{1}(x),\Psi{}_{M}^{3}(y)\right\} \\
\left\{ \Psi{}_{M}^{2}(x),\Phi{}_{E}^{1}(y)\right\}  & \left\{ \Psi{}_{M}^{2}(x),\Phi{}_{E}^{2}(y)\right\}  & \left\{ \Psi{}_{M}^{2}(x),\Phi{}_{E}^{3}(y)\right\}  & \left\{ \Psi{}_{M}^{2}(x),\Psi{}_{M}^{1}(y)\right\}  & \left\{ \Psi{}_{M}^{2}(x),\Psi{}_{M}^{2}(y)\right\}  & \left\{ \Psi{}_{M}^{2}(x),\Psi{}_{M}^{3}(y)\right\} \\
\left\{ \Psi{}_{M}^{3}(x),\Phi{}_{E}^{1}(y)\right\}  & \left\{ \Psi{}_{M}^{3}(x),\Phi{}_{E}^{2}(y)\right\}  & \left\{ \Psi{}_{M}^{3}(x),\Phi{}_{E}^{3}(y)\right\}  & \left\{ \Psi{}_{M}^{3}(x),\Psi{}_{M}^{1}(y)\right\}  & \left\{ \Psi{}_{M}^{3}(x),\Psi{}_{M}^{2}(y)\right\}  & \left\{ \Psi{}_{M}^{3}(x),\Psi{}_{M}^{3}(y)\right\}
\end{array}\right),
\]
}{\scriptsize \par}

\[
=\left(\begin{array}{cccccc}
0 & 0 & 0 & -1 & 0 & 0\\
0 & 0 & 0 & 0 & -1 & 0\\
0 & 0 & 0 & 0 & 0 & -1\\
1 & 0 & 0 & 0 & 0 & 0\\
0 & 1 & 0 & 0 & 0 & 0\\
0 & 0 & 1 & 0 & 0 & 0
\end{array}\right)\frac{\delta(x-y)}{4\pi c},
\]
this matrix has rank=6  and zero null vectors, therefore this theory has not secondary constraints. Because of we do not expect  secondary  constraints for this theory, the evolution in time  of the primary constraints   will allow  us to know the six Lagrange multiplies introduced  above, this is  

\begin{eqnarray}
\dot{\Phi}_{E}^{n}(x)&=&\int d^{3}y\left\{ \Phi{}_{E}^{n}(x),\mathcal{H}_{c}(y)\right\} +\int d^{3}yu_{m}(y)\left\{ \Phi{}_{E}^{n}(x),\Phi{}_{M}^{m}(y)\right\}, \\\nonumber
&=&\frac{1}{8\pi}\epsilon_{ijn}\partial_{j}E_{i}(x)+\frac{1}{8\pi}\epsilon_{ijn}\partial_{j}E_{i}(x)-\frac{1}{4\pi c}u_{n}(x)\approx0,\\\nonumber
\end{eqnarray}
hence, 
\begin{equation}
u_{i}(x)=c\epsilon_{kji}\partial_{j}E_{k}(x)=-c\epsilon{_{i}}^{{jk}}\partial_{j}E_{k}(x),
\label{eq.10a}
\end{equation}
and 
\begin{eqnarray}
\dot{\Psi}{}_{M}^{m}(x) &=& \int d^{3}y\left\{ \Psi{}_{M}^{m}(x),\mathcal{H}_{c}(y)\right\} +\int d^{3}yv_{n}(y)\left\{ \Psi{}_{M}^{m}(x),\Phi{}_{E}^{n}(y)\right\}, \\\nonumber
&=&\frac{1}{8\pi}\epsilon_{ijm}\partial_{j}B_{i}(x)+\frac{1}{8\pi}\epsilon_{ijm}\partial_{j}B_{i}(x)+\frac{1}{4\pi c}v_{m}(x)\approx0,\\\nonumber
\end{eqnarray}
then 
\begin{equation}
v_{i}(x)=-c\epsilon_{kji}\partial_{j}B_{k}(x)=c\epsilon{_{i}}^{jk}\partial_{j}B_{k}(x),
\label{eq.12a}
\end{equation}
thus,  the six Lagrange multipliers have  been identified. \\
We have observed that  the complete set  of  constraints are given by (\ref{eq.8}),  which  are of second class. In fact,  we  see  that 
\begin{eqnarray}
\left\{ \Phi{}_{E}^{i}(x),\Phi{}_{E}^{j}(y)\right\} &=&0, \nonumber \\
\left\{ \Phi{}_{M}^{i}(x),\Psi{}_{M}^{j}(y)\right\} &=&0, \nonumber \\
\left\{ \Phi{}_{E}^{i}(x),\Psi{}_{M}^{j}(y)\right\} &=&-\frac{1}{4\pi c}\delta^{ij}\delta(x-y).
\end{eqnarray}
With all those results at hand,  we are able to calculate the physical degrees of freedom as follows; there are 6 dynamical variables and  6 second class constraints, thus, there are three  physical degrees of freedom. However, we need to take into account  the equations (\ref{eq.3}). For this aim, we observe that the constraints (\ref{eq.8}) satisfy  the following two reducibility conditions 
\begin{eqnarray}
\partial_i \Phi^{i}_E=0, \nonumber \\
\partial_i \Psi^i_M =0.
\end{eqnarray}
Therefore,   reducibility conditions imply  that there are $[6-2]=4$ second class constraints, hence the  physical degrees of freedom are four. This result is expected because the action (\ref{eq.1}) is not invariant under parity, therefore the degrees of freedom are distinguishable under parity, however, we know that in the case of Maxwell theory, the action is invariant under parity and the degrees of freedom are not distinguishable under parity.  
It is important to remark,  that in spite of action  (\ref{eq.1}) yields Maxwell equations of motion, our results indicate that the action (\ref{eq.1}) does not describe the dynamics  of the electromagnetic field at all. In fact, it is well-know that Maxwell theory is a gauge theory and  has two physical degrees of freedom; on the other side,  Eq. (\ref{eq.1}) is not a gauge theory and has four physical degrees of freedom. In this manner,  we confirm that a dynamical system should be defined by means an action principle plus  their equations of motion,  and not only by means the equations of motion. Our results obtained in this letter,  show  relevant differences among the action (\ref{eq.1})  and conventional Maxwell action at classical level  and of course,   will be interesting research  the quantum differences among them as well, for this aim  we will develop all the necessary tools in follow sections.   \\
Because of there are  second class constraints in the  theory, we shall calculate the Dirac's brackets.  Dirac's brackets will be useful in order to  study  the  observables,  as well as, for performing the  quantization of the theory.  Hence, Dirac's brackets are defined by  
\[
\{F(x),G(y)\}_{D}\equiv\{F(x),G(y)\}+\int d^{3}zd^{3}w\{F(x),\chi^{\alpha}(z)\}W_{\alpha\beta}^{-1}\{\chi^{\beta}(w),G(y)\},
\]
where  $W_{\alpha\beta}^{-1}$ is the  inverse of the matrix $W$ defined above,  and $\chi^\alpha= (\Phi{}_{E}^{j},\Psi{}_{M}^{i})$ are the second class constraints. Hence, we obtain  the following Dirac's brackets 
\begin{eqnarray}
\{E_{i}(x),\Pi_{E}^{j}(y)\}_{D}&=&\{E_{i}(x),\Pi_{E}^{j}(y)\}+\int d^{3}zd^{3}w\{E_{i}(x),\chi^{\alpha}(z)\}W_{\alpha\beta}^{-1}\{\chi^{\beta}(w),\Pi_{E}^{j}(y)\} \nonumber \\
&=&\{E_{i}(x),\Pi_{E}^{j}(y)\} \nonumber \\
&=&\frac{1}{3}\delta_{i}^{j}\delta(x-y).\\\nonumber
\end{eqnarray}
\begin{eqnarray}
\{E_{i}(x),B^{j}(y)\}_{D}&=&\{E_{i}(x),B^{j}(y)\}+\int d^{3}zd^{3}w\{E_{i}(x),\chi^{\alpha}(z)\}W_{\alpha\beta}^{-1}\{\chi^{\beta}(w),B^{j}(y)\} \nonumber \\
&=&4\pi c\delta_{i}^{j}\delta(x-z).\\\nonumber
\end{eqnarray}
\begin{eqnarray}
\{\partial_{i}E^{i}(x),B^{j}(y)\}_{D}&=&\{\partial_{xi}E_{i}(x),B^{j}(y)\}+\int d^{3}zd^{3}w\{\partial_{xi}E_{i}(x),\chi^{\alpha}(z)\}W_{\alpha\beta}^{-1}\{\chi^{\beta}(w),B^{j}(y)\} \nonumber \\
&=&4\pi c\partial_{j}\delta(x-y).\\\nonumber
\end{eqnarray}
\begin{eqnarray}
\{B_{i}(x),P_{M}^{j}(y)\}_{D}&=&\{B_{i}(x),P_{M}^{j}(y)\}+\int d^{3}zd^{3}w\{B_{i}(x),\chi^{\alpha}(z)\}W_{\alpha\beta}^{-1}\{\chi^{\beta}(w),P_{M}^{j}(y)\} \nonumber \\
&=&\{B_{i}(x),P_{M}^{j}(y)\} \nonumber \\
&=&\frac{1}{3}\delta_{i}^{j}\delta(x-y).
\end{eqnarray}

\begin{eqnarray}
\{\partial_{i}\Pi_{E}^{i}(x),B_{j}(y)\}_{D}&=&\{\partial_{i}\Pi_{E}^{i}(x),B_{j}(y)\}+\{\partial_{i}\Pi_{E}^{i}(x),\chi^{\alpha}(z)\}W_{\alpha\beta}^{-1}\{\chi^{\beta}(w),B_{j}(y)_{j}\} \nonumber \\
&=&\partial_{i}\{\Pi_{E}^{i}(x),B_{j}(y)\}+\{\partial_{i}\Pi_{E}^{i}(x),\Psi{}_{M}^{i}(z)\}W_{ij}^{-1}\{\Psi{}_{M}^{j}(w),B_{j}(y)_{j}\}=0. 
\end{eqnarray}

\begin{eqnarray}
\{\partial_{i}P_{M}^{i}(x),E_{j}(y)\}_{D}&=&\{\partial_{i}P_{M}^{i}(x),E_{j}(y)\}+\{\partial_{i}P_{M}^{i}(x),\chi^{\alpha}(z)\}W_{\alpha\beta}^{-1}\{\chi^{\beta}(w),E_{j}(y)_{j}\} \nonumber \\
&=&\partial_{i}\{P_{M}^{i}(x),E_{j}(y)\}+\{\partial_{i}\Pi_{E}^{i}(x),\Phi{}_{E}^{i}(z)\}W_{ij}^{-1}\{\Phi{}_{E}^{j}(w),E_{j}(y)_{j}\}=0. 
\end{eqnarray}
We finish our  analysis by calculating the extended action and the extended Hamiltonian. For this aim, we use the Lagrange multipliers found in (\ref{eq.10a})  and (\ref{eq.12a}), and we find 
\begin{eqnarray}
S_{E}[A_{\mu},\pi^{\mu},\lambda_{j},u_{i}] &=&\int d^{4}x[\Pi_{E}^{i}\dot{E}_{i}+P_{M}^{i}\dot{B}_{i}-\frac{1}{8\pi}[B_{i}\epsilon_{ijk}\partial_{j}B_{k}+E_{i}\epsilon_{ijk}\partial_{j}E_{k}]  \nonumber \\
&-& u_{i}\Psi^{i}-v_{i}\Phi^{i}-\overline{u}_{i}\Psi^{i}-\overline{v}_{i}\Phi^{i}] \nonumber \\
&=&\int d^{4}x[\Pi_{E}^{i}\dot{E}_{i}+P_{M}^{i}\dot{B}_{i}-[c\Pi_{E}^{i}\epsilon_{ijk}\partial_{j}B_{k}-cP_{M}^{i}\epsilon_{ijk}\partial_{j}E_{k}]-\overline{u}_{i}\Psi^{i}-\overline{v}_{i}\Phi^{i}] 
\end{eqnarray}
where we can identify the extended Hamiltonian given by 
\begin{equation}
H_{E}=\int d^{3}x \left(c\Pi_{E}^{i}\epsilon_{ijk}\partial_{j}B_{k}-cP_{M}^{i}\epsilon_{ijk}\partial_{j}E_{k} \right).
\label{eq.22}
\end{equation}
It is easy to see that $H_E$ is of first class. In fact,  we have 
\begin{eqnarray}
\{ \Phi^i_E, H_E \} &=& c\epsilon^{ij}{_{l}} \partial_j \Psi^l \approx 0, \nonumber \\
\{\Psi^i_M , H_E \}&=&   -c  \epsilon^{ij}{_{l}} \partial_j \Phi^l \approx 0.
\end{eqnarray}
On the other hand,  by using the equations (15)-(20) we observe that  the Dirac's  bracket among $H_E$ and the second class constraints vanish, therefore $H_E$ is an observable. \\
From the extended  action we calculate the following variations 
\begin{eqnarray}
\delta S_{E}&=&\int d^{4}x[\delta\Pi_{E}^{i}\dot{E}_{i}+\Pi_{E}^{i}\delta\dot{E}_{i}+P_{M}^{i}\delta\dot{B}_{i}+\delta P_{M}^{i}\delta\dot{B}_{i}-c\delta\Pi_{E}^{i}\epsilon_{ijk}\partial_{j}B_{k}-c\Pi_{E}^{i}\epsilon_{ijk}\partial_{j}\delta B_{k}] \nonumber \\
&+&c\delta P_{M}^{i}\epsilon_{ijk}\partial_{j}E_{k}+cP_{M}^{i}\epsilon_{ijk}\partial_{j}\delta E_{k}-\delta\overline{u}_{i}\Psi^{i}-\delta\overline{v}_{i}\Phi^{i}],\nonumber \\
&=&\int d^{4}x[\delta\Pi_{E}^{i}\partial_{t}E_{i}+\partial_{t}(\Pi_{E}^{i}\delta E_{i})-\delta E_{i}\partial_{t}\Pi_{E}^{i}+\partial_{t}(P_{M}^{i}\delta B_{i})-\delta B_{i}\partial_{t}P_{M}^{i}+\delta P_{M}^{i}\partial_{t}B_{i}\nonumber \\
&-&c\delta\Pi_{E}^{i}\epsilon_{ijk}\partial_{j}B_{k}]-c\epsilon_{ijk}\partial_{j}(\Pi_{E}^{i}\delta B_{k})+c\epsilon_{ijk}\delta B_{k}\partial_{j}\Pi_{E}^{i}+c\delta P_{M}^{i}\epsilon_{ijk}\partial_{j}E_{k}+c\epsilon_{ijk}\partial_{j}(\delta P_{M}^{i}E_{k})\nonumber \\
&-&c\epsilon_{ijk}\delta E_{k}\partial_{j}P_{M}^{i}-\delta\overline{u}_{i}\Psi^{i}-\delta\overline{v}_{i}\Phi^{i}]\nonumber \\
&=&\int d^{4}x[(\partial_{t}E_{i}-c\epsilon_{ijk}\partial_{j}B_{k})\delta\Pi_{E}^{i}-(\partial_{t}\Pi_{E}^{i}+c\epsilon_{kji}\partial_{j}P_{M}^{k})\delta E_{i}-(\partial_{t}P_{M}^{i}-c\epsilon_{kji}\partial_{j}\Pi_{E}^{k})\delta B_{i}]\nonumber \\
&+&(\partial_{t}B_{i}+c\epsilon_{ijk}\partial_{j}E_{k})\delta P_{M}^{i}-\delta\overline{u}_{i}\Psi^{i}-\delta\overline{v}_{i}\Phi^{i}],\\\nonumber
\end{eqnarray}
where the following equations of motion rise 
\begin{eqnarray}
\delta E_{i}\;:\;\partial_{t}\Pi_{E}^{i} &=&c\epsilon_{ijk}\partial_{j}P_{M}^{k}, \nonumber \\
\delta B_{i}\;:\;\partial_{t}P_{M}^{i}&=&-c\epsilon_{ijk}\partial_{j}\Pi_{E}^{k},\nonumber \\
\delta\Pi_{E}^{i}\;:\;\partial_{t}E_{i}&=&c\epsilon_{ijk}\partial_{j}B_{k}, \nonumber \\
\delta P_{M}^{i}\;:\;\partial_{t}B_{i}&=&-c\epsilon_{ijk}\partial_{j}E_{k}, \nonumber \\
\delta\overline{u}_{i}\;:\;\Psi^{i}&\approx&0, \nonumber \\
\delta\overline{v}_{i}\;:\;\Phi^{i}&\approx&0.
\end{eqnarray}
It is important  to remark, that the above equations of motion  were not obtained in \cite{7,8}. In fact, in those works it  was ignored  that Eq. (\ref{eq.1}) is a singular system and  it was defined a dynamical system by using the equations of motion, then, the system was treated as a non-singular system, however, we have showed that Eq. (\ref{eq.1}) does not describes Maxwell theory. On the other hand, the Hamiltonian found in \cite{7,8} is not equivalent to Eq. (\ref{eq.22}), the Hamiltonian found in this work  by following Dirac's method is of first class, and the  dynamics of the system is carry out on the constraints surface defined on the full phase space. Nevertheless, in \cite{7,8} is not possible to  talk about   first class or second class constraints  and the dynamics of the system is carry out  on the full phase space, but we have showed that this is not possible because Eq. (\ref{eq.1}) is  singular. 
\section{A quantum state of zero energy}
In order to observe differences among the Eq.(1) and conventional Maxwell theory at quantum level, we will calculate a quantum state of zero energy for the Hamiltonian  (\ref{eq.22}). For this aim, we identify the following classical-quantum correspondence for the canonical momentum $\Pi^i_E \rightarrow-i \frac{\delta }{\delta E_i}$ and  $P^i_M \rightarrow -i \frac{\delta}{ \delta B_i}$ for the magnetic field. Hence
\begin{eqnarray}
\left( \widehat{E}_i(x) \psi \left(E,B \right)  \right) &=&   \widehat{E}_i \psi(E,B), \quad \quad \quad \quad \left( \widehat{B}_i(x) \psi \left(E,B \right)  \right) =   \widehat{B}_i \psi(E,B),  \nonumber \\ 
\left( \widehat{\Pi}^i_E(x)  \psi \left(E,B \right)  \right) &=& -i \frac{\delta \psi(E,B)}{\delta E_i}, \quad \quad \quad \quad \left( \widehat{P}^i_M(x)  \psi \left(E,B \right)  \right) = -i \frac{\delta \psi(E,B)  }{\delta B_i},
\end{eqnarray}
where $\psi(E,B)$ is an arbitrary function of the fields $E$, $ B$  and represents a quantum state. By using the above  correspondence,  the classical-quantum representation of the Hamiltonian (\ref{eq.22})  is given by 
\begin{eqnarray}
\widehat{H}= \int \left( -i c \epsilon_{i}{^{jk}} \partial_jB_k \frac{\delta}{\delta E_i} + ic \epsilon_i {^{jk}}\partial_jE_k \frac{\delta}{ \delta B_i}   \right), 
\end{eqnarray}
the representation of the vacuum for the theory will correspond to an eigenfunction of zero energy for the Hamiltonian $\widehat{H}$ determined by 
\begin{equation}
\widehat{H} \psi (E,B)=0.
\label{eq.28}
\end{equation}
Hence, the function that solves exactly Eq. (\ref{eq.28}) is  given by 
\begin{equation}
\psi(E,B)= e^{\alpha I(E,B)},
\label{eq.29} 
\end{equation}
where $\alpha$ is  a constant and
\begin{equation}
I(E,B)= \frac{1}{2} \int \left(  \epsilon^{ijk} E_i \partial_j E_k +  \epsilon^{ijk} B_i \partial_j B_k  \right) dx^3.
\label{eq.30}
\end{equation} 
Some remarks are important to comment;   the wave function given in (\ref{eq.29}) does not correspond  the Chern-Simons state known for Maxwell theory, because the dynamical variable is  the connexion. In fact, it is well-known that in Maxwell theory,  the wave function  that solves the Hamiltonian  for the vacuum is given for the Chern-Simos state for the connexion $A$, therefore, with this result we confirm  that  at quantum level,   the action given in Eq.(\ref{eq.1}) does not describes  the electromagnetic field. Furthermore, the expression (\ref{eq.30}) can be written as 
\begin{equation}
I(E,B)= \frac{1}{4} \int  \left(  \epsilon^{ijk} E_i {\mathbf{R_{jk}}} +  \epsilon^{ijk} B_i {\mathbf{\Upsilon _{jk} }}\right), 
\end{equation} 
where ${\mathbf{R_{jk}}}= \partial _i E_j - \partial_j E_i$ and ${\mathbf{\Upsilon _{jk} }} = \partial _i B_j - \partial_j B_i$ are some like curvatures of the field $E$ and $B$ respectively. Hence, it is straightforward to show that   $I(E,B)$ is invariant under changing 
\begin{eqnarray}
E_i \rightarrow E_i &+& \partial_i \theta, \nonumber \\
B_i \rightarrow B_i &+& \partial_i \theta, \nonumber \\
\end{eqnarray}
therefore,  $I(E,B)$ is  a composition of  Chern-Simons terms associated for the fields $E$ and $B$. Finally, it is easy to show that Eq.(\ref{eq.30}) viewed as a field theory is topological one and diffeomorphism covariant, thus, while in  the action (\ref{eq.1}) the fields $E$ and $B$ are not  gauge fields, for $I(E,B)$ they are. In this manner, the state (\ref{eq.29}) is not physically accepted for the theory. Of course, will be interesting to  perform the path integral quantization of the action   (\ref{eq.29})  in order to obtain a better understanding of the quantization of the system. All these ideas are in progress and will be reported in forthcoming works. 
\section{Concluding remarks}
In this paper, we have performed a complete Dirac's analysis for an action yielding Maxwell's equations of motion. In our analysis we found that although   the non conventional action yields the same equations of motion than Maxwell theory,  its symmetries   are not those  associated to the electromagnetic field. The theory studied  is not a gauge theory and does not have  the same number of degrees of freedom. In this manner, we need to be careful for  defining  a dynamical  system. If we  define a system by using only the equations of motion, we are in the situation that  an infinity number of Hamiltonian structures can bee defined for the same system, however, this fact presents a  problem because by  changing   the hamiltonian structure,   we could obtain  several actions  yielding the same equations of motion, but the symmetries of the theory  can be lost,  such as  it was presented in the analysis  developed  along this paper. Therefore, in order to know  the symmetries, a dynamical system must be defined by means of an action principle,  because the action contains the relevant  information and symmetries of the theory. \\
It is important to remark  that similar results will be found if the canonical analysis is performed for the case of an action close to linearized  gravity. In fact, in \cite{7, 8} is proposed an action with similar structure than (\ref{eq.1})
\begin{equation}
S[E_{ij}, B_{ij}]= \int \left[B^{ij}\left (\frac{1}{c}\partial_{t}E_{ij}-\epsilon{_{i}}^{kl}\partial_{k}B_{lj}\right)-E^{ij} \left(\frac{1}{c}\partial_{t}B_{ij}+\epsilon{_{i}}^{kl}\partial_{k}E_{lj} \right)\right]d^{4}x,
\label{eq.24}
\end{equation} 
where $E_{ij}$ and $B_{ij}$  are related with the components of the curvature tensor corresponding to the perturbed metric (see \cite{8,7} for full details).  However, from the analysis performed in this paper, we will found that the action (\ref{eq.24}) does not describe the dynamics of the linearized  gravity theory;  the principal symmetry  as gauge invariance  is lost  and the number of physical degrees of freedom do not correspond to  those found in  linearized gravity \cite{9}. Finally, in \cite{10} can be found an alternative hamiltonian description of gravity, however, in  that work the analysis was performed in the same way  than  \cite{7, 8}, this is,    defining a dynamical system  by means its equations of motion, hence the symmetries well-knew  in gravity are lost. 
\newline
\newline
\newline
\noindent \textbf{Acknowledgements}\\[1ex]
This work was supported by Sistema Nacional de Investigadores M\'exico. The authors  want   to thank to R. Cartas-Fuentevilla  and J. J. Toscano for useful conversations on the subject.

\end{document}